\documentclass[a4paper,twoside]{article}
\usepackage{placeins}
\usepackage{epsfig}
\usepackage{subfigure}
\usepackage{calc}
\usepackage{amssymb}
\usepackage{amstext}
\usepackage{amsmath}
\usepackage{amsthm}
\usepackage{multicol}
\usepackage{pslatex}
\usepackage{apalike}
\usepackage{color}
\usepackage{listings}
\usepackage{tikz,ifthen}
\usepackage{hyperref}
\usepackage{SCITEPRESS}     

\usetikzlibrary{arrows,trees,backgrounds,automata,shapes,decorations,plotmarks,fit,calc,positioning,shadows,chains,
  patterns,decorations.pathreplacing}
\tikzstyle{block}=[rectangle, draw, thin, inner sep=3pt, text centered, drop shadow, fill=orange!20!yellow!20]
\tikzstyle{pre}=[<-,shorten <=1pt,>=stealth']
\tikzstyle{post}=[->,shorten >=1pt,>=stealth']
\tikzstyle{bi}=[<->,shorten >=1pt,shorten <=1pt,>=stealth']
\tikzstyle{every initial by arrow}=[initial text={},initial distance=1em,post]
\tikzstyle{every state}=[minimum size=0.4cm,drop shadow,fill=orange!20!yellow!20]
\tikzstyle{transition}= [post,shorten >=1pt,node distance=2cm, inner sep=2pt,bend angle=20]
\usetikzlibrary{calc}

\DeclareFixedFont{\ttb}{T1}{txtt}{bx}{n}{7.5} 
\DeclareFixedFont{\ttm}{T1}{txtt}{m}{n}{7.5}  

\definecolor{deepblue}{rgb}{0,0,0.5}
\definecolor{deepred}{rgb}{0.6,0,0}
\definecolor{deepgreen}{rgb}{0,0.5,0}

\newcommand{\ra}{\rangle}

\newcommand{\compose}{\parallel}

\newcommand{\request}{{\color{blue}request}}
\newcommand{\waitfor}{{\color{green!50!black}wait for}}
\newcommand{\blocking}{{\color{red}blocking}}

\renewcommand\j[1]{\textsc{#1}}

\begin{document}

\title{On-the-Fly Construction of Composite Events \\ in Scenario-Based Modeling Using Constraint Solvers}

\author{}

\author{\authorname{Guy Katz\sup{1}, Assaf Marron\sup{2},   Aviran Sadon\sup{3}, and Gera Weiss\sup{3}}
\affiliation{The Hebrew University of Jerusalem, Jerusalem, Israel}
\affiliation{Weizmann Institute of Science, Rehovot, Israel}
\affiliation{ Ben-Gurion University of the Negev, Be'er Sheva, Israel}
\email{guykatz@cs.huji.ac.il, assaf.marron@weizmann.ac.il,
sadonav@post.bgu.ac.il,
geraw@cs.bgu.ac.il}
}


\abstract{
Scenario-Based Programming is a methodology for modeling and
constructing complex reactive systems from simple, stand-alone
building blocks, called scenarios. These scenarios are designed to
model different traits of the system, 
and can be interwoven together and executed to
produce cohesive system behavior. Existing execution frameworks
for scenario-based programs
allow scenarios to 
specify their view of what the system must, may, or must not do
only through very strict interfaces. 
This limits the methodology's expressive power and often prevents
users from modeling certain complex 
requirements.
Here, we propose to extend
Scenario-Based Programming's execution mechanism to 
allow scenarios to specify how the system should behave 
using rich logical constraints. We then leverage modern
constraint solvers (such as SAT or SMT solvers) to 
resolve these constraints 
at every step of running the system,
towards yielding the desired overall system behavior. 
We provide an implementation of our
approach and demonstrate its applicability to various systems that
could not be easily
modeled in an executable manner
by existing Scenario-Based approaches.
}

\onecolumn \maketitle \normalsize \vfill

\section{\uppercase{Introduction}}
\label{sec:introduction}
\noindent

Modeling complex systems is a difficult and error-prone task. 
The emerging \emph{Scenario-Based Programming} approach (\emph{SBP})~\cite{DaHa01,HaMa03,HaMaWeACM} aims
 to mitigate this difficulty. 
The key notion underlying SBP is modeling 
through the specification
of \emph{scenarios}, each of which represents a certain
 aspect of the system's behavior. These scenarios may describe either
desirable behaviors that the system should strive to uphold, or
undesirable behaviors that the system should try to avoid. The models
produced in SBP are fully executable:
when composed together according
to certain underlying semantics, the scenarios yield cohesive system
behavior.

The SBP approach has been implemented
in dedicated frameworks such as the Play-Engine and PlayGo for the visual language of Live Sequence Charts (LSC) language~\cite{HaMa03,Playgo2010} or
ScenarioTools~\cite{GeGrGuKoGlMaKa17} for the Scenario Modeling Langauge (SML) textual language. Further, SBP has been
implemented on top of several standard programming 
languages, such as Java~\cite{HaMaWe10BPJ}, C++~\cite{HaKa14}, and JavsScript~\cite{BaWeRe18}, and was amalgamated with the Statecharts visual formalism~\cite{morse2018SCSBP}

SBP has been successfully used in modeling complex systems,
such as web-servers~\cite{HaKa14}, cache coherence
protocols~\cite{HaKaMaMa16}, robotic controllers~\cite{GrGr18b}, and as part of the \emph{Wise Computing} effort aimed at turning computers into proactive members of system development teams~\cite{HaKaMaMa18IEEEComputer}.

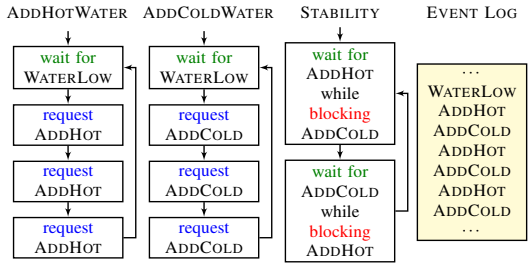
\begin{figure}[htp]
  \centering
  \scalebox{0.65} {
    
    \tikzstyle{box}=[draw,  text width=2cm,text centered,inner sep=3]
    \tikzstyle{set}=[text centered, text width = 10em]

    \begin{tikzpicture}[thick,auto,>=latex',line/.style ={draw, thick, -latex', shorten >=0pt}]
      
      \matrix(bts) [row sep=0.3cm,column sep=2cm]  {

        \node (box1)  [box] {\waitfor{} \j{WaterLow}}; \\
        \node (box2)  [box] {\request\ \j{AddHot}}; \\
        \node (box3)  [box] {\request\ \j{AddHot}}; \\ 
        \node (box4)  [box] {\request\ \j{AddHot}}; \\ 
      };

      \draw [->] ($(box1.north) + (0,0.3cm)$) -- (box1.north);
      \node (title) [above=0.1cm of bts,box,draw=none] at ($(bts) + (-0.25cm,2.51cm)$) 
      {\j{AddHotWater}};  
      
      \begin{scope}[every path/.style=line]
        \path (box1)   -- (box2);
        \path (box2)   -- (box3);
        \path (box3)   -- (box4);
        \path (box4.east)   -- +(.25,0) |- (box1);
      \end{scope}

      \matrix(bts2) [right=.25cm of bts, row sep=0.3cm,column sep=2cm] {
        \node (box1)  [box] {\waitfor{} \j{WaterLow}}; \\
        \node (box2)  [box] {\request\ \j{AddCold}}; \\
        \node (box3)  [box] {\request\ \j{AddCold}}; \\ 
        \node (box4)  [box] {\request\ \j{AddCold}}; \\ 
      };
      
      \draw [->] ($(box1.north) + (0,0.3cm)$) -- (box1.north);
      \node (title) [above=0.1cm of bts2,box,draw=none] at ($(bts2) + (-0.25cm,2.51cm)$) 
      {\j{AddColdWater}};

      \begin{scope}[every path/.style=line]
        \path (box1)   -- (box2);
        \path (box2)   -- (box3);
        \path (box3)   -- (box4);
        \path (box4.east)   -- +(.25,0) |- (box1);
      \end{scope}

      \matrix(bts3) [right=.25cm of bts2, row sep=0.3cm,column sep=2cm] {
        \node (box1)  [box] {\waitfor{}  \j{AddHot} while  \blocking\ \j{AddCold}}; \\
        \node (box2)  [box] {\waitfor{}  \j{AddCold} while \blocking\ \j{AddHot}}; \\
      };

      \draw [->] ($(box1.north) + (0,0.3cm)$) -- (box1.north);
      \node (title) at (title-|bts3) [box,draw=none] {\j{Stability}};  

      \begin{scope}[every path/.style=line]
        \path (box1)   -- (box2);
        \path (box2.east)   -- +(.25,0) |- (box1);
      \end{scope}
      
      \node (log)  [right=.3cm of bts3,box,text width=2cm,fill=yellow!20] {
        $\cdots$ \\ 
        \j{WaterLow} \\
        \j{AddHot}  \\ 
        \j{AddCold} \\ 
        \j{AddHot}  \\ 
        \j{AddCold} \\ 
        \j{AddHot}  \\ 
        \j{AddCold} \\ 
        $\cdots$ \\
      }; 

      \node (title2) at (title-|log)            
      [box,draw=none] {\j{Event Log}};  
    \end{tikzpicture}
  }  
  \caption{
    (From~\cite{HaKaMaWe14})
    A scenario-based model of a
    system that controls the water level in a tank with hot and
    cold water taps. Each scenario object is depicted as a transition
    system, where the nodes represent the predetermined synchronization points.
    The scenario object
    \j{AddHotWater} repeatedly waits for \j{WaterLow} events and requests three
    times the event \j{AddHot}; and the 
    scenario object \j{AddColdWater} performs a symmetrical operation
    with cold water. In a model that includes only the objects \j{AddHotWater} and
    \j{AddColdWater}, the three \j{AddHot} events and three
    \j{AddCold} events may be triggered in any order during execution.
    In order to maintain the stability of the water temperature in the tank,
    the scenario object
    \j{Stability} enforces the interleaving of \j{AddHot} and \j{AddCold} events 
    by using event blocking. The execution trace of the resulting model appears
    in the event log.
  }  
  \label{fig:watertap}
\end{figure}

Despite the diversified adaptations of SBP for various programming
languages and for various use cases, a common theme remains: in all
existing frameworks, scenarios are interwoven using a very basic mechanism.
Specifically, during execution the scenarios are
synchronized at predetermined points, and at every synchronization point each
scenario declares a set of events it would like to see triggered, and a
set of events it forbids from being triggered. The execution framework
then selects for triggering one event that is requested by at least one scenario and is not 
blocked by any of the scenarios. The event is broadcast to all scenarios, and the execution 
continues until the next synchronization point is reached. An example appears in Fig.~\ref{fig:watertap}.

It has been suggested that some of the benefits of SBP
come from this basic event selection semantics. In particular, the event selection
mechanism is sufficiently simple to make scenario-based models easy to analyze
automatically using formal compositional  techniques~\cite{HaKaKaMaMiWe13,HaKaMaWe15,HaKaMaMa16,GrGr18}, and even to automatically distribute, repair and synthesize them~\cite{HaKaMaWe12,HaKaKamaWeWi15,StGrGrHaKaMa17,GrGrKaMaGlGuKo16,GrGrKaMa16}, primarily because it facilitates the automatic 
composition of individual scenarios that are 
simple and succinct~\cite{HaKaLaMaWe15}.
Still, the simplicity of the
event selection mechanism seems to be a limiting
factor in some cases --- requiring cumbersome workarounds to associate complex behaviors with simple events, and at times even preventing the use of SBP for modeling a particular system altogether.

Consider, as a toy example, a model for an autonomous drone. The model
 contains various behavioral scenarios for
modeling the drone's horizontal and vertical movement. 
At every execution cycle of the model, independent actions
may be triggered for each of the axes ---  \emph{climb}, \emph{descend}, or \emph{maintain
height} for the vertical axis, and \emph{turn right}, \emph{turn left} or \emph{maintain
direction} for the horizontal axis.
Further, climb or descend actions
are parameterized by a numerical value indicating the angular velocity of
the climb or descent; and similarly, turn right or turn left actions
are parameterized by the angular velocity of the turn. It is unclear how to express
such a model in SBP. For example, because the traditional event
selection mechanisms stipulates that precisely one action be triggered
in every cycle, how shall we express the fact that
multiple actions (horizontal and vertical) may be triggered in the same 
cycle? And how shall we account for the infinitely-many numerical parameters for
ascent, decent and turning actions? Some discretization schemes may be
proposed, but this seems to go against the grain of SBP ---
which aims at creating simple and intuitive scenario objects.

In this paper, we propose an extension to SBP that utilizes \emph{constraint
solvers}: automated tools that take as input a set of variables and
certain kinds of constraints on these variables, and
return a variable assignment that satisfies the given
constraints (or indicate that no such assignment exists). Automated solvers have become widespread and highly
successful in the last decades, particularly in tasks related to program
analysis and verification~\cite{ClHeVeBl18}. Here, we propose to use
such solvers \emph{on-the-fly, as
part of the execution mechanism} of scenario-based
models. Specifically, we propose to augment SBP such that in each
synchronization point, each scenario contributes to the creation of a
formula that is fed 
into the constraint solver --- and the assignment (of all variables) which is returned by the
solver assumes the role of the event selected for triggering.
Further, the very act of selecting simple events from some set is extended into constructing, or computing complex events based on rich specifications. 
This allows us to specify scenarios that interact using a far richer
formalism, and can thus model more complex systems. Compared to
existing SBP approaches, this allows for the scenarios to
collaboratively construct the events, not only choose among events
that each propose.

In particular,
using constraint solvers in this fashion allows us to seamlessly model the
autonomous drone system: the constraints produced in every
synchronization point may include multiple variables indicating
multiple actions; and these constraints may include arbitrary
numerical values, indicating, e.g., the various angular velocity parameters. We
elaborate on this example later on. 

In this work we describe how a solver-based SBP modeling framework can be implemented, focusing mainly on the semantics but also propose a syntax, with accompanying implementation details and examples for completeness.

The paper is organized as follows. In Section~\ref{sec:background} we provide some necessary background on SBP and on constraint solvers. In Section~\ref{sec:sbpWithConstraints} we propose our extension to SBP that allows modelers to integrate it with constraints solvers, followed by illustrative examples. In Section~\ref{sec:evaluation} we describe an evaluation of our approach, followed by a discussion of related work in Section~\ref{sec:relatedWork}.  We conclude in Section~\ref{sec:conclusion}.

\section{\uppercase{Background}}
\label{sec:background}
\noindent

\subsection{Scenario-Based Modeling}
Before we discuss our proposed extensions to SBP, we begin by
recapping the existing, commonly used formulation and semantics.
Formally, a scenario-based model consists of independent scenario
 objects that are interwoven at run time. Each scenario
repeatedly declares sets of events which, from its own perspective, should,
may, or must not occur. At runtime, the scenarios are executed 
simultaneously and are synchronized by a mechanism responsible
for selecting events that constitute the integrated system
behavior. The scenarios never interact with each other directly; all
interactions are carried out through the event selection mechanism.

Following the definitions in~\cite{Ka13}, we define a scenario object $O$ over event set $E$
as the tuple 
 $O = \langle Q, \delta, q_0, R, B \ra$, where the components are
 interpreted as follows:
\begin{itemize}
\item $Q$ is a set of states, each representing one of the
  predetermined synchronization points;
\item $q_0$ is the initial state;
\item $R:Q\to 2^E$ and $B:Q\to 2^E$
map states to the sets of events requested and blocked at these
states (respectively); and
\item $\delta: Q \times E \to 2^Q$ is a transition function,
  indicating how the object reacts when an event is triggered.
  \end{itemize}

Scenario objects can be composed, in the following manner.
For objects
 $O^1 = \langle Q^1, \delta^1, q_0^1, R^1, B^1 \ra$ and
 $O^2 = \langle Q^2, \delta^2, q_0^2, R^2, B^2 \ra$ over a
 common event set $E$, the composite scenario object $O^1\compose O^2$ is defined by
$O^1 \compose O^2 = \langle Q^1 \times Q^2, \delta,
\langle q_0^1,q_0^2\rangle, R^1\cup R^2, B^1\cup B^2\rangle
$,
where:
\begin{itemize}
\item $\langle \tilde{q}^1,\tilde{q}^2\rangle \in  \delta(\langle q^1,q^2\rangle, e)$
if and only if $\tilde{q}^1 \in \delta^1(q^1,e)$ and $\tilde{q}^2\in
\delta^2(q^2,e)$; and
\item The union of the labeling functions is defined in the natural way; e.g. $e\in (R^1\cup
R^2)(\langle q^1,q^2 \rangle)$ if and only if $e \in R^1(q^1) \cup
R^2(q^2)$, and
$e\in (B^1\cup
B^2)(\langle q^1,q^2 \rangle)$ if and only if $e \in B^1(q^1) \cup
B^2(q^2)$.
\end{itemize}

A \emph{behavioral model} $M$ is simply a collection of scenario
objects  $O^1, O^2,\ldots, O^n$,
and the executions of $M$ are the executions of the composite object
$O = O^1\compose O^2\compose\ldots\compose O^n$.
Each such execution starts from the initial state of $O$,
and in each state $q$ along the run an enabled event is chosen for triggering, if
one exists (i.e., an event $e\in R(q) - B(q)$).
 Then, the execution moves to state $\tilde{q}\in \delta(q,e)$, and
so on. 

\subsection{Constraint Solvers}
As our proposed extensions to SBP rely heavily on automated constraint
solvers, we give here a very brief introduction to some of these
tools (and mention sources of information for additional reading).
Broadly speaking, constraint solvers are automated tools that take as input a set of  
constraints given as a formula $\varphi$ over a set of variables $V$, and either (i) return a variable 
assignment that satisfies $\varphi$, or (ii) answer that no such variable assignment exists. 
(A satisfying assignment is usually called a \emph{model}, but we will refrain from using that term as to not
overload it). Different solvers differ in the kinds of constraints they allow as part of their input, and many popular solvers
operate on constraints given in restricted forms of first order logic. The performance of these solvers (and the complexity of the
problems they solve) also closely depends on the inputs they allow.

In this paper, we will focus on three kinds of automated solvers:

\textbf{Boolean Satisfiability (SAT) Solvers.} These are solvers
  that operate on a set $V$ of Boolean variables, and limit the
  constraint formula $\varphi$ to be a quantifier-free propositional
  formula over the variables of $V$. The solver then attempts to find 
  a Boolean assignment that satisfies $\varphi$.
  For example, for $V=\{p,q\}$, the formula $\varphi_1=(p\vee q)\wedge(p\vee \neg q)$ is
  satisfiable, and one satisfying assignment is $p,\neg q$; whereas the
  formula $\varphi_2=(\neg p\vee \neg q)\wedge p\wedge q$ is
  unsatisfiable. Although the Boolean satisfiability problem is
  NP-complete, there exist many mature tools that can solve instances
  with hundreds of thousands of variables~\cite{Nadel2009}. A particular kind of SAT solvers, called \emph{MaxSAT} solvers, attempt to find a Boolean assignment that satisfies as many of the input constraints as possible (and not necessarily all of the constraints).

\textbf{Linear Programming (LP) Solvers.} LP solvers operate
  on a set $V$ of rational variables, and the constraint formula
  $\varphi$ is a conjunction of linear constraints, often referred to
  as a \emph{linear program}.
  For example, for the variables $V=\{x,y,z\}$, the constraint $\varphi_3=(x\leq 5) \wedge
  (x+y\leq z)$ is satisfiable, whereas the constraint $\varphi_4=(x\leq 5)
  \wedge (y\leq 2) \wedge (x+y\geq 20)$ is unsatisfiable. LP is known to be solvable in polynomial time, although
  many solvers use worst-case exponential algorithms that turn out to
  be more efficient in practice~\cite{Chvatal1987}.
 
\textbf{Satisfiability Modulo Theories (SMT) Solvers.}
These
  solvers can be regarded as generalized SAT solvers, capable of
  handling formulas in rich fragments of first order logic.
  The satisfiability of the formulas is checked
  modulo \emph{background theories}, which intuitively restrict the search
  only to satisfying assignments that ``make sense'' according to these
  certain theories. For example, considering the theory of arrays of
  integer elements with variable set $V=\{a,b\}$, the formula
  $\varphi_5=(a[3]\geq b[5])\wedge (a[4]\leq b[0])$ is satisfiable,
  whereas the formula $\varphi_6=(a=b)\wedge(a[4]\neq b[4])$ is unsatisfiable.
  Modern SMT solvers support many theories of interest, including
  various arithmetic theories, the theory of uninterpreted functions,
  and theories of arrays, of sets, of strings, and
  others~\cite{BaTi18}. Further, these background theories can be combined: for example, one can define formulas that includes arrays of integers or sets of strings, etc. The SMT problem is, in general, undecidable,
  although certain background theories afford efficient decision procedures.

The three kinds of solvers are used for different tasks, and
all are highly successful. Many mature tools exist, and a great deal
of research is being put into improving them further.

\section{\uppercase{Integrating SBP with
    Constraint Solvers}}
\label{sec:sbpWithConstraints}
\noindent

\subsection{Extending SBP}
The notion underlying our proposed extension of SBP is as follows. At each synchronization point, instead of declaring sets of requested and blocked events, each scenario object $O_i$ can instead declare a set of constraint formulas $\Phi=\{\varphi_i^1,\ldots,\varphi_i^l\}$ that are intended as guiding rules for a solver-based mechanisms that assembles the events. 
These constraint formulas are \emph{labeled} by a labeling function $L_i$, which takes a formula $\varphi_i^k$ and returns its labeling, i.e.  a subset of a finite set of predefined labels $\mathcal{L}$. 
The motivation for these labels is that they can be used to assign different semantics to different constraint formulas.

For example, going back to the drone system described in the
introduction, one scenario can specify that the total speeds of the
rotors must be above some threshold and another scenario can suggest
to increase one of the rotors. The labeling function is a protocol
through which the execution mechanism knows that the first is a
``must'' specification and the latter is a ``may'' condition. 

At each synchronization point, the execution mechanism collects the sets of constraint formulas $\Phi_1,\ldots,\Phi_n$ produced by the individual scenario objects, and combines them into a global constraint formula $\varphi$. This formula is then passed into a constraint solver, and the satisfying assignment returned by the solver is broadcast to all scenarios, which can then change their states. If no satisfying assignment is found, the SBP model is deadlocked, and the execution terminates. (Another possible extension in case a deadlock is discovered is to wait for an external event, along the lines proposed in~\cite{HaMaWeWi11}, but this is beyond our scope here).

Formally, we modify the definitions of SBP to support integration with constraint solvers as follows. Let $V$ denote a set of variables,
and let $\mathcal{L}$ denote a finite set of labels.
We define a scenario object $O$ over $V,\mathcal{L}$ as a tuple $O = \langle Q, \delta, q_0, C, L \rangle$ , where $Q$ is a set of states and $q_0$ is the initial state, as before. The function $C$, which replaces the
 labeling functions $R$ and $B$ in the previous definition, takes a state $q\in Q$ as input and returns a set of constraint formulas $\Phi=\{\varphi^1,\ldots,\varphi^l\}$ over the
 variables of $V$. The function $L$ returns a labeling of these constraint formulas according to the current state, i.e. $L:Q \times \xi\rightarrow2^\mathcal{L}$, where $\xi$ represents the set of all possible formulas. By convention, we require that $L(q,\varphi)=\emptyset$ for every $\varphi$ such that $\varphi\notin C(q)$.
 The transition function $\delta$ is now defined as $\delta: Q\times A(V)\to 2^Q$, where $A(V)$ is the set of all possible assignments to the variables of $V$.
 Intuitively, given a specific state $q$ and a variable assignment $\alpha \in A(V)$, invoking $\delta(q,\alpha)$ returns the set of states the object may transition into.

In order to account for the new constraint formulas, we modify the composition operator for scenario objects as follows:
For objects $O^1 = \langle Q^1, \delta^1, q_0^1, C^1, L^1 \rangle$ and $O^2 = \langle Q^2, \delta^2, q_0^2, C^2, L^2 \rangle$ over a common variable set $V$ and a common label set $\mathcal{L}$, the composite scenario object $O^1\compose O^2$ is defined by
$O^1 \compose O^2 = \langle Q^1 \times Q^2, \delta,
\langle q_0^1,q_0^2\rangle, C, L\rangle
$, where $\langle \tilde{q}^1,\tilde{q}^2\rangle \in  \delta(\langle q^1,q^2\rangle, \alpha)$  if and only if $\tilde{q}^1 \in \delta^1(q^1,\alpha)$ and $\tilde{q}^2\in \delta^2(q^2,\alpha)$.
The constraint-generating function $C$ is defined as $C(\langle q^1,q^2\rangle)=C^1(q^1)\cup C^2(q^2)$, i.e. the constraints defined by the individual objects are combined and become the constraints defined by the composite object. We define $L(\langle q^1,q^2\rangle,\varphi)= L^1(q^1,\varphi) \cup L^2(q^2,\varphi)$ using again the convention that $L^i(q^i,\varphi) = \emptyset$ if $\varphi\notin C^i(q^i)$.

The key difference between our extended semantics and the original is in the event selection mechanism. As before, a \emph{behavioral model} $M$ is a collection of scenario objects  $O^1, O^2,\ldots, O^n$, and the executions of $M$ are the executions of the composite object $O = O^1\compose O^2\compose\ldots\compose O^n$.
Each such execution starts from the initial state of $O$, and after each state $q$ along the run a variable assignment $\alpha$ is assembled by invoking a constraint solver on a formula $\varphi$ constructed from $C(q)$, according to the constraint labeling $L$. Specifically, we assume that the modeler also provides a \emph{constraint composition rule} $\psi$. Given the constraint-generating function $C$ and the labeling function $L$, $\psi$ dictates how to construct for every state $q$ the constraint formula $\varphi$ that should be passed to the solver, and/or how to treat the various constraints altogether (e.g., apply priorities among scenarios, or apply various optimization goals when multiple solutions exist). 
The execution then moves to state $\tilde{q}\in \delta(q,\alpha)$, and
so on.  

\subsection{Illustrative Examples}

The aforementioned framework is general, and can be customized in several ways through the constraint formulas, their labeling, and the constraint composition rule $\psi$. We next illustrate this using a few simple examples.

\textbf{Traditional SBP Semantics.}
The traditional semantics of SBP can be obtained as follows. We allow only two labels $\mathcal{L}=\{r,b\}$, where $r$ represents \emph{request constraints} and $b$ represents \emph{block constraints}. In addition, we define the variable set $V$ to contain precisely one variable, $V=\{e\}$, representing the triggered event. Next, we syntactically restrict the constraint formulas
$\varphi_i$ to be of the form $e=c$ for some constant $c$; and finally, for any state $q$ we define the constraint composition rule to be:
\[
\psi(q,C,L)=(\bigvee_{\varphi\in C(q)\ |\ r \in L(q,\varphi)}\varphi)\wedge(\bigwedge_{\varphi\in C(q)\ |\ b \in L(q,\varphi)}\neg\varphi)
\]
Intuitively, at each state, each scenario object can declare events it would like to see triggered (expressed as constraints labeled $r$), and those it wants to prevent from being triggered (expressed as constraints labeled $b$). The constraint composition rule then translates these individual constraints into a global formula representing the fact that the triggered event needs to be requested and not blocked (note that constraints labeled $b$ are negated).

When using these particular restrictions, the straightforward
solver of choice is a SAT solver: since the formula $\varphi$ only contains propositional connectives and the variable $e$ can only take on a finite number of values, we can encode these possible values using a finite set of Boolean variables (this process is often called \emph{bit-blasting}). A modern SAT solver can then be used for selecting the triggered event very quickly --- in a way that is likely to enable an execution that is sufficiently fast for many application domains.

\textbf{Autonomous Drone.}
The general framework we proposed in the previous subsection can be
used to model   
complex interactions, which are either beyond the reach of the
traditional semantics, or at least require a great deal of
effort on the modeler's side. Let us return to our toy aircraft example: a drone capable of
simultaneous vertical and horizontal maneuvers. Using our extended
modeling framework, we can define our variable set $V$ to include two
variables, $V=\{v,h\}$, where $v$ represents the vertical angular velocity 
and $h$ represents the horizontal angular velocity. One scenario object can be used for
setting upper and lower bounds on the vertical turning angular velocities, due to the drone's
mechanical limitations (see Fig.~\ref{fig:verticalLimit}), and another can be used for limiting the horizontal turning angular velocity (see Fig.~\ref{fig:horizontalLimit}). In this case we require no labeling of the constraint, i.e. $\mathcal{L}=\emptyset$, and the constraint composition rule $\psi$ is a simply a conjunction of all the individual constraints.

\begin{figure}[htp]
  \centering
  \scalebox{0.65} {
    
    \tikzstyle{box}=[draw,  text width=3cm,text centered,inner sep=3]
    \tikzstyle{set}=[text centered, text width = 10em]

    \begin{tikzpicture}[thick,auto,>=latex',line/.style ={draw, thick, -latex', shorten >=0pt}]

      \node (box1)  [initial,box] {$\varphi_1= -5\leq v\leq 5$};

      \path[] (box1) edge [loop below,thick] node {true} (box1);
    \end{tikzpicture}
  }  
  \caption{A scenario object that puts hard limits on the
    vertical turning angular velocity of the drone. The scenario has a single
    synchronization point (indicated by a single state), in which it
    contributes $\varphi_1= -5\leq v\leq 5$ to the global constraint set. The only
    transition, a self loop that does not depend on the variable assignment returned by the solver, indicates that the scenario continues to
    contribute this constraint, regardless of the satisfying
    assignment discovered by the solver.}
  \label{fig:verticalLimit}
\end{figure}

\begin{figure}[ht]
  \centering
  \scalebox{0.65} {
    
    \tikzstyle{box}=[draw,  text width=3cm,text centered,inner sep=3]
    \tikzstyle{set}=[text centered, text width = 10em]

    \begin{tikzpicture}[thick,auto,>=latex',line/.style ={draw, thick, -latex', shorten >=0pt}]

      \node (box1)  [initial,box] {$\varphi_2= -10\leq h\leq 10$};

      \path[] (box1) edge [loop below,thick] node {true} (box1);
    \end{tikzpicture}
  }  
\caption{A scenario object that puts hard limits on the
  horizontal turning angular velocity of the drone.}
  \label{fig:horizontalLimit}
\end{figure}

Without any additional limitations, i.e. if only these two scenarios
existed in the system, the constraint formula in any synchronization
point would be $\varphi=\varphi_1\wedge\varphi_2 = (-5\leq v\leq 5)\wedge(-10\leq h\leq 10)$.
Because the constraint are arithmetical, linear constraints, we can use an LP solver to dispatch them; and indeed, in this case an LP solver
will return an assignment
such as $v=3, h=0$. Other objects in the system, called actuators, may
then process these values and adjust the drone's engines accordingly.

Let us now consider a particular flight situation. Suppose another object is in charge of navigating the drone to
its destination, and that that object is requesting a right turn at an angular velocity of at least 6 degrees per second:
 $\varphi_3 = h\geq 6$. Further suppose that a sensor has
detected an electrical wire up ahead, and in order to circumvent it is
requesting either that the elevation be increased, or that a left turn
be initiated: $\varphi_4= h \leq -3 \vee v \geq 2$. In that case, when
the solver is given the global constraint formula
$\varphi=\wedge_{i=1}^4\varphi_i$, a possible solution is $h=8,v=3$
--- which satisfies all constraints, by both turning right and
increasing the drone's altitude.

\textbf{Dependency Management.}
So far, we have seen two examples for constraint composition rules
$\psi$: when simulating the traditional SBP event selection mechanism,
we labeled individual constraints as request or block statements, and
then composed them accordingly; and in the drone example, we had no
labeling, and $\psi$ was a simple conjunction. We now demonstrate a
situation in which yet another composition rule is useful.

Consider a system in charge of installing software packages on a
computer, similar to the standard package managers that ship with
modern Linux distributions. Software packages have
\emph{dependencies}: for example, 
installing package A might require that package B already be installed, in which case we say that package A \emph{requires} package B. Some packages may also be \emph{incompatible} with other packages: for example, if package A is incompatible with package C, this means that A cannot be installed alongside C.
The \emph{state} of the system is the set of currently 
installed software packages. Finally, the system is given a
user-supplied goal, such as ``install A''. In order to achieve the goal, the system needs to
install A and any required packages, while removing the smallest
number of packages currently installed that A and its dependencies are incompatible with. Of course, deciding which packages to install and which to remove in order to achieve an optimal result is a complex task.

To model this system using our extended version of SBP, we can utilize a specific kind of SAT
solver, called a \emph{MaxSAT} solver. A MaxSAT input formula 
consists of subformulas labeled either \emph{hard} or
\emph{soft}, and the solver finds an assignment that
satisfies the hard constraints, and as many of the soft constraints as
possible. MaxSAT solvers play a crucial role in our model, in the following way:
for each package dependency, we will introduce a scenario object that adds a hard constraint that represents the dependency; and we will introduce other scenario objects that express the currently-installed packages as soft
constraints. That way, the MaxSAT solver will give us back an assignment that indicates which packages should be installed and which should be removed, in a way that guarantees that the goal package is installed while the number of previously installed packages that need to be removed is minimized~\cite{MaBoDiVoDuLe06,ArLy08}.

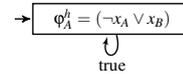
\begin{figure}[ht]
 \centering
  \scalebox{0.65} {
    
    \tikzstyle{box}=[draw,  text width=3cm,text centered,inner sep=3]
    \tikzstyle{set}=[text centered, text width = 10em]

    \begin{tikzpicture}[thick,auto,>=latex',line/.style ={draw, thick, -latex', shorten >=0pt}]

      \node (box1)  [box,initial] {$\varphi^h_A= (\neg x_A\vee x_B)$};

      \path[] (box1) edge [loop below,thick] node {true} (box1);
    \end{tikzpicture}
  }  
\caption{A scenario object that encodes the fact the
  installing A requires B. Observe that the constraint is labeled as hard, to indicate that it must never be violated.}
  \label{fig:dependencyA}
\end{figure}

\begin{figure}[ht]
\centering
  \scalebox{0.65} {
    
    \tikzstyle{box}=[draw,  text width=3cm,text centered,inner sep=3]
    \tikzstyle{set}=[text centered, text width = 10em]

    \begin{tikzpicture}[thick,auto,>=latex',line/.style ={draw, thick, -latex', shorten >=0pt}]

      \node (box1)  [initial,box] {$\varphi^s_B= x_B$};

    \node (box2)  [box, right = 3cm of box1] {$\varphi = \emptyset$};

      \path[->] (box1) edge [loop below,thick] node {$x_B$} (box1);
      \path[->] (box1) edge [bend right,thick] node {$\neg x_B$} (box2);
      
      \path[->] (box2) edge [loop below,thick] node {$\neg x_B$} (box2);
      \path[->] (box2) edge [bend right,thick] node[swap] {$x_B$} (box1);
    \end{tikzpicture}
  }  
\caption{A scenario object that adds $x_B$ as a soft
  constraint if package $B$ is currently installed (left state), and contributes no constraints if it is not installed (right state). Switching between the states is performed according to the assignment discovered by the solver --- specifically, it depends on whether $x_B$ is assigned to true or not. We assume the package is initially installed.}
  \label{fig:dependencyB}
\end{figure}
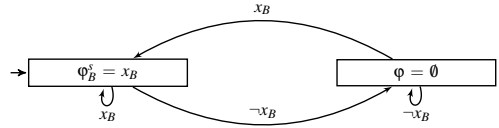

More specifically, our model for the package dependency system is constructed as follows. The
variable set $V$ consists of a Boolean variable for each software package,
e.g. $\{x_A,x_B,x_C,\ldots\}$, that signifies whether the package is
installed (variable is true) or not installed (variable is false). A
change in the variable's value indicates that the package should be
installed or removed. Our label set is $\mathcal=\{h,s\}$, indicating whether a constraint is hard or soft, respectively.
Each dependency is represented by a dedicated
object; for example, the requirement ``A requires B'' is encoded by the scenario object in Fig.~\ref{fig:dependencyA}. Other objects are used for encoding the soft constraints representing the currently
installed packages---an example appears in Fig.~\ref{fig:dependencyB}.

To avoid clutter, we omit the scenario object in charge of reading the installation goal from the user, and the scenario objects of the actuators in charge of monitoring changes in consecutive variable assignments and translating these changes into the installation or removal of packages.

\section{\uppercase{Implementation and Evaluation}}
\label{sec:evaluation}
\noindent
In this section we evaluate the applicability of our approach by discussing its implementation, and by applying it to more complex problems.

\subsection{Two Implementations} 
We developed a proof-of-concept implementation of our approach in two platforms. 
The first uses MATLAB/Simulink. Scenario objects generate their constraints as strings. These strings are then passed into MATLAB \texttt{solve}, the equation and system solver. The solution yielded by the solver is then translated into variable values that control classical Simulink-driven behavior. The results of this behavior are also fed back into the scenarios, which can then change the constraints they present. 

Below, we describe in detail a second implementation, based on 
Python and the Z3 SMT solver~\cite{DeBj08Z3}. The framework enables users to create fully-executable models using the aforementioned approach, and then run them and analyze the output. We plan to make the framework available online in the near future, and also intend to extend it; see some discussion in Section~\ref{sec:conclusion}.

We began by implementing the basic SBP semantics in our framework. For these semantics, the 
set of allowed labels is $\mathcal{L}=\{\text{\texttt{may,must,wait-for}}\}$: the \texttt{may} label represents requested events, the \texttt{must} label is used
here to block the complement of the specified event set, 
and the \texttt{wait-for} label is merely syntactic sugar used to simplify defining the transition relation. This labeling scheme uses the composition rule
\begin{align*}
  \psi(q,C&,L)=\\
  &(\bigvee_{\varphi\in C(q)\ |\ \text{\texttt{may}} \in
    L(q,\varphi)}\varphi)\wedge(\bigwedge_{\varphi\in C(q)\ |\ \text{\texttt{must}} \in
    L(q,\varphi)}\varphi)
\end{align*}
For the event selection mechanism, we apply the Z3 solver for
solving the formula $\varphi$, constructed from the scenario objects' \texttt{may} and \texttt{must} constraints as specified above.

In our implementation, each scenario object is modeled using a Python \emph{generator}:
a function that can pause itself and yield control at any point, and then be subsequently resumed when it is re-invoked with the language's \texttt{next()} idiom. This functionality of Python allows us to implement the SBP idioms --- i.e., have the scenario objects pause at synchronization points and be resumed when a satisfying assignment for the variables of $V$ has been found.

At each synchronization point, the scenario object thus yields control, and passes to the event selection
mechanism a Python \emph{dictionary} containing any subset of the keys \texttt{may}, \texttt{must}, and
\texttt{wait-for}, where each such key is associated with a Z3 constraint.

 The core of the code of  the execution mechanism appears in Fig.~\ref{fig:pythonImplementation}.
 The main function, \emph{run}, takes as input the set of scenario objects, and then executes the model that is obtained by composing these objects. Specifically, the function invokes the scenario objects, one at a time, and waits for each of them to reach its next synchronization point, indicated by a \texttt{yield} statement.  Once all scenario objects are synchronized, the framework     collects the constraints (in the form of dictionaries, called \emph{tickets} in the code)  generated by
 the individual scenarios.  These constraints are then composed and passed on to Z3, which tries to find an assignment that satisfies all the \texttt{must} and \texttt{may} constraints.  If such an assignment is found, the execution framework wakes up the scenario objects whose \texttt{wait-for} conditions are  satisfied by the chosen assignment, and allows them to resume. They then continue to execute until they reach the next \texttt{yield} point, and then the process is repeated again, possibly ad infinitum.

\begin{figure*}[htp]
\begin{lstlisting}[label={code:execution},frame = single,tabsize=2]
def run(scenarios):
    global m      # A variable where the solved model is published
    tickets = []  # A list containing the tickets issued by the scenarios

    # Run all scenario objects to their initial yield
    for sc in scenarios:
        ticket = next(sc)       # Run the scenario to its first yield and collect the ticket
        ticket['sc'] = sc       # Maintain a pointer to the scenario in the ticket
        tickets.append(ticket)  # Add the ticket to the list of tickets

    # Main loop
    while True:
        # Compute a disjunction of may constraints and a conjunction of must constraints
        (may, must) = (False, True)
        
        for ticket in tickets:
            if 'may' in ticket:
                may = Or(may, ticket['may'])    
            if 'must' in ticket:
                must = And(must, ticket['must']) 

        # Compute a satisfying assignment and break if it does not exist
        sl = Solver()
        sl.add(And(may, must))
        if sl.check() == sat:
            m = sl.model()
        else:
            break  

        # Reset the list of tickets before rebuilding it
        oldTickets = tickets
        tickets = []
    
        # Run the scenarios to their next yield and collect new tickets 
        for oldTicket in oldTickets:
            # Check whether the scenario waited for the computed assignment
            if 'wait-for' in oldTicket and is_true(m.eval(oldTicket['wait-for'])):
                
                # Run the scenario to the next yield and collect its new ticket 
                newTicket = next(oldTicket['sc'], 'ended') 
    
                # Add the new ticket to the list of tickets (if the scenario didn't end)
                if not newTicket == 'ended':
                    newTicket['sc'] = oldTicket['sc'] # Copy the pointer to the scenario 
                    tickets.append(newTicket)
            else:
                # Copy the old tickets to the new list
                tickets.append(oldTicket)
    
            
\end{lstlisting}
  \caption{A python implementation of an extended SBP framework. The
    scenarios are assumed to be given as an array of generators that
    return, using the yield command, labeled Z3 propositions expressed
    as dictionaries with keys that are subsets of \{\texttt{may},
    \texttt{must}, \texttt{wait-for}\}. In this code these
    dictionaries are called ``tickets''.}
  \label{fig:pythonImplementation}
  \end{figure*}

\subsection{Examples}
\textbf{Hot-Cold example.}
Using this framework, one can specify the scenario objects from the
water tank system that appears in Fig.~\ref{fig:watertap}. This
specification appears in Fig.~\ref{fig:waterTapImplementation}. When
the scenario objects defined therein are executed, the satisfying
assignments obtained by the solver during the execution alternate
between assigning ``hot'' to true and ``cold'' to false, and vice
versa. 

\begin{figure}[ht]
\begin{lstlisting}[frame = single]
hot = Bool('hot')
cold = Bool('cold')

def mutual_exclusion():
	yield {'must': Or(Not(hot),Not(cold))}

def three_hot():
	for i in range(3):
		yield {'may': hot, 'wait-for': hot}

def three_cold():
	for j in range(3):
		yield {'may': cold, 'wait-for': cold}

def no_two_same_in_a_row():
  yield {'wait-for': true}
	while True:
	  if is_true(m[cold]):
	 	  yield {'must':Not(cold),'wait-for':true}
	  if is_true(m[hot]):
		  yield {'must':Not(hot),'wait-for':true}
\end{lstlisting}
      \caption{A simple example of a model that uses the solver-based execution mechanism. The model sets the ``hot'' and ``cold'' flags, indicating that additional doses of hot and cold water are added to a tank, according to the following five rules:  (1) do not add hot and cold doses at the same time; (2) add three doses of hot water; (3) add three doses of cold water; (4) never add two doses of the same type in a row. }
      \label{fig:waterTapImplementation}
      \end{figure}
      
\begin{figure}[ht]
	\begin{lstlisting}[frame = single]
temp = Real('temp')

def hot_temp():
	yield {'must': Implies(hot, temp > 50)}

def cold_temp():
	yield {'must': Implies(cold, temp < 50)}	

def after_hot_temp():
	while True:
		yield {'wait-for': hot}
		while is_true(m[hot]):
			yield {'must':temp>20,'wait-for':true}

def after_cold_temp():
	while True:
		yield {'wait-for': cold}
		while is_true(m[cold]):
      yield {'must': temp<80,'wait-for':true}
	\end{lstlisting}
	\caption{New requirements for the water tap model: (1) the temperature of a hot event must be above 50; (2) the temperature of a cold event must be below 50; (3) the temperature of an event that follows a hot event must be above 20; (4) the temperature of an event that follows a cold event must be below 80.}
	\label{fig:waterTapNewRequirements}
\end{figure}

Consider now a situation where the customer decides to change the requirements for the system. For example, assume that the last requirement (that does not allow to add two doses of the same type in a row) is removed and, instead, the customer decides to add the requirements modeled in Fig.~\ref{fig:waterTapNewRequirements}. The scenarios listed in the figure are then added instead of the last scenario in Fig.~\ref{fig:waterTapImplementation}.

Note that the new requirements involve a new solver variable called
``temp'', for temperature, that the new scenarios control. Note also
that this is done without changing anything in the remaining scenarios
and that the remaining scenarios are not at all aware of the new
variable. 

This example raises the following discussion: consider, for example, the situation in Fig.~\ref{fig:maymustUnintendedBehaviour} where, as in the water tap example, two scenarios deal with separate variables called $x_1$ and $x_2$, respectively. Since the first scenario is not aware of the second one, it assumes that the only \texttt{may} constraint for $x_1$ is that it is greater than $50$ --- and so it does not expect the solver to allow an assignment to $x_1$ that is smaller than $50$. According to our semantics, however, the composition rules produces the constraint $\psi=x_1>50 \vee x_2>50$ to which the assignment $\{x_1= 0, x_2 = 51\}$ is valid. A way to avoid this unintended behavior can be to label each proposition with the variable that it is aware of and to solve for each set of variables separately. Another way to avoid it can be to look for assignments that maximize the number of satisfied \texttt{may} constraints, e.g by using solvers that optimize the number of satisfied clauses.

\begin{figure}[ht]
\begin{lstlisting}[frame = single,mathescape=true]
def scenario1():
	yield {'may': $x_1 > 50$}

def scenario2():
	yield {'may': $x_2 > 50$}
\end{lstlisting}
	\caption{An example of an unintended behavior with the \texttt{may} and \texttt{must} semantics. Since $\psi=(x_1 {<}50) \vee (x_2{>}50)$, the assignment $\{x_1=0, x_2=51\}$
	is valid despite the fact that no scenario specified that $x_1$ may be smaller than 50. To fix this issue, we propose to change the composition rule so that each set of variables and constraints is solved separately. }
	\label{fig:maymustUnintendedBehaviour}
\end{figure}

\textbf{Leader follower benchmark example.}
As a more complex example, we used the extended SBP modeling framework, with the composition rule described in preceding sub-section, to model a reactive controller for a rover in a \emph{leader-follower simulation}.
In a leader-follower system, a controlled follower rover tracks a leader rover. The follower rover is required to follow the leader, while always staying at a safe distance from it, no matter how the leader behaves (assuming reasonable bounds on speed and turn angles). This problem served as a challenge problem in the MDETOOLS'18 workshop, where the organizers supplied a simulation software for it. Participants of the workshop were encouraged to demonstrate their various modeling approaches by 
constructing software to control the follower rover (see \url{mdetools.github.io/mdetools18/challengeproblem.html}). 

The simulator provided in the MDETOOLS'18 challenge periodically emits the location of the rovers, the distance between the rovers, and the heading angle of the follower (compass). The follower rover can be controlled by setting the power for the left and right wheels in the range $\{-100,\dots,100\}$. For example, if power to the left wheels is set to $40$ and power to the right wheels is set to $0$, the rover will turn right.

The code for the scenarios that we created in order to control the follower rover is listed in Fig.\ref{fig:wLF}. The first scenario specifies the bounds for the $pR$ and $pL$ variables, indicating the power to the right and left wheels, respectively. The second scenario specifies forward and backward motion, where wheel power is a function of the relative distance, i.e., when the rovers get too far apart or too close, the follower gradually increases or decreases power to the wheels, even down to negative values. The third scenario specifies how the follower is steered towards the leader location. When the relative angle (calculated from the data emitted from the simulator) exceeds a specified value (3 degrees), the follower will accordingly turn left or right towards the leader. The last scenario specifies how to perform a turn by setting different power levels to the left and right wheels (note, however, that this scenario does not trigger a turn --- but rather controls a turn that has been triggered by another scenario).
This example demonstrates the modularity of the suggested approach and the ability to construct complex behaviors using distinct behavioral aspects.

The final behavior yielded in this case study is 
indeed very similar to the one yielded by the traditional behavioral
programming approach where events are selected without a constraint
solver, using direct filtering logic, which had been presented in~\cite{GrBaWeSaMa18}. The main 
difference between the techniques used in these two implementations is that in the
implementation described in~\cite{GrBaWeSaMa18} scenarios can only
request finite sets of events while here the \lstinline{spin()}
scenario, for example, specifies infinitely many options that may
happen. This allows, as demonstrated by the \lstinline{turnpowers()}
scenario, to break the specification to better align with the
requirements. 

\begin{figure*}[ht]
\begin{lstlisting}[frame=single]
def bounds():
	yield {'must': $-MAX \leq pL \leq MAX \wedge -MAX \leq pR \leq MAX$}
	
def forward_backward():
	while True:
		if dist > CLOSE:
		  if dist < FAR:
		     yield {'may': $pL=pR=\frac{MAX(dist-CLOSE)}{FAR - CLOSE}$, 'wait-for': true}
		  else:
		     yield {'may': $pL=pR=MAX$, 'wait-for': true}
	         
		else:
		  if dist > VERY_CLOSE:
		    yield {'may': $pL = pR = \frac{MAX(dist-CLOSE)}{CLOSE -VERY\_CLOSE}$, 'wait-for': true}
		  else:
		    yield {'may':  $pL=pR=-MAX$, 'wait-for': true}
            
def spin():
	while True:
		if abs(dir_error) > 3:
			if dir_error > 0:
			  yield {'may': $pL>pR$, 'must': $pL>pR$,  'wait-for': true}
			else:
			  yield {'may': $pL<pR$, 'must': $pL<pR$, 'wait-for': true}
		else:
			yield {'wait-for': true}

def turnpowers():
	yield {'must': $pL \neq pR \Rightarrow (pL=0 \wedge pR=40) \vee (pL=40 \wedge pR=0)$}
	   
\end{lstlisting}
      \caption{Main scenarios of the leader-follower model. The first
        scenario specifies the bounds for the $pR$ and $pL$ variables,
        which represent the power for the left  and  right follower
        wheels. The second specifies the follower forward and
        backwards motion as a function of the distance from the
        leader. The third  specifies how the follower is steered
        towards the leader location as function of $dir\_error$, which
        represents the relative angle (in degrees). The last scenario specifies the turn powers.}
      \label{fig:wLF}
\end{figure*}

\textbf{A Patrol Vehicle.}
Another example, described briefly to fit space constraints, was implemented in MATLAB/Simulink and associated solvers with the tool described earlier in this section. It is a simulation of an autonomous vehicle that moves repeatedly in a fixed route in the shape a figure eight.
The main scenarios reflect the following requirements: (1) The vehicle should always attempt to accelerate to a maximum prespecified speed; (2) when the vehicle reaches a sharp curve, it should reduce its speed below a specified value until exiting the curve; and (3) after driving at a speed that is higher than a certain value, for a length of time that is higher than some threshold, 
the vehicle must reduce its allowed speed and acceleration to some other values for a certain amount of time  (e.g., to avoid engine overheating). 

This example illustrates and emphasizes the power of scenarios as
``stories'' that progress from one state to another and present different constraints at different times and states. E.g., specifying the speed constraints that hold only after detecting the arrival at (or departure from) a sharp curve, or the passage of a certain amount of time, appears more intuitive,  and is better aligned with the stated requirements, than specifying ever-present constraints with conjunctions of conditions, of, say, current speed and road curvature, or, current speed and acceleration and the time that has passed since certain events in the past. 

\section{\uppercase{Related Work}}
\label{sec:relatedWork}
\noindent

The paper presents a particular approach to run-time composition of behavior, namely, extending the existing SBP-style composition with specification and solving of constraints. Below we briefly compare SBP to other execution-time composition mechanism with a special focus  on the present context of constraint specifications (see~\cite{HaMaWeACM} for an earlier, related analysis). 

A key contribution of SBP over most other approaches to system specification is its succinctness and intuitiveness. These properties emerge from the ability to specify forbidden behavior explicitly and directly, rather than as control-flow conditions that prevent certain pieces of code or specification from actually doing the undesired action
(this was accomplished first with concrete lists of requested events and filter-based blocking, and now, more generally, with constraint solvers).
For example, in SBP, one can build, and sometimes even test, the specification that a vehicle is not allowed to enter a road intersection when the traffic light is red, before having coded how vehicles behave.  
By comparison, other approaches, like business-workflow engines, simulation engines, and tools for test-driven development support intuitive specification of executable use cases and scenarios, but their support for generic composition of multiple scenarios and anti-scenarios is limited.
Ordinary procedural and object oriented programming,
functional programming and logic programming languages provide for composition of behaviors, 
but the requirements’ scenarios and use cases are not directly visible in the code and are reflected only in emergent properties of the actual execution.

SBP principles have been implemented in several languages in both distributed and centralized environments. These implementations also position  SBP as a design pattern for using common constructs like semaphores, messaging, and threads, as well as concepts such as agent-orientation for incrementally and alignment of code with requirements.

Publish-subscribe mechanisms provide for straightforward parallel composition, but without language support for forbidden behavior. Aspect oriented programming (AOP)~\cite{Kiczales1997AspectOriented} supports specifying and executing cross-cutting concerns on top of a base application, but does not support specifying  forbidden behavior, state management within an aspect, or symmetry between aspects and base code, which SBP does.

Behavior-based models such as Brooks's subsumption architecture~\cite{Brooks1986Subsumption} Branicky's behavioral programming~\cite{branicky1999behavioral}, and LEGO Mindstorms leJOS (see review in~\cite{arkin1998behavior}), also call for constructing systems from behaviors.
SBP is a language-independent formalism with multiple implementations and extends in a variety of ways each of the coordination and arbitration mechanisms in those architectures.

The execution semantics of behavioral programming has similarities to the event-based scheduling of SystemC~\cite{IEEE2006SystemC}, which performs cyclical co-routine scheduling  by synchronization, evaluation, update and notification. SBP differs from SystemC in 
its direct support for specifying scenarios and anti-scenarios with direct relation to the original  requirements, where SystemC provides a particular architecture for composing parallel component in certain architectures and designs.
In SBP the synchronization is an inherent technique for  continuously complying with all  constraints that the requirements impose where in SystemC synchronization is used for coordination in an otherwise parallel component execution. This also implies differences in the details in the semantics of synchronization, event selection, queuing, and state management within a parallel component. 
 
The BIP language (behavior, interaction, priority) and the concept of glue for assembling components~\cite{Sifakis2008ConcurBIP} pursue goals similar to SBP’s with a focus on correctness by construction rather than on execution of intuitively specified behaviors and constraints, with run-time resolution of these constraints. 

As mentioned earlier, SBP was recently implemented in the visual formalism of Statecharts. The Yakindu Statecharts tool extended Statecharts' original support for orthogonal, concurrent and hierarchical state machines~\cite{Harel87Statecharts}, with optional specification of requested and blocked events in any state, and a corresponding enhancement to the event selection semantics~\cite{morse2018SCSBP}. These enhancements also provide the formal definitions of SBP principles, which are based on state machines and transition systems (see,  e.g.,~\cite{HaMaWe10BPJ}), with a direct, concrete, executable implementation that is also readily understood by humans.
This facilitates direct casting of inter-object behaviour, which usually is only emergent when modeling with statecharts and other state-machine languages, in the same language and formalism as intra-object behaviour. 

In SBP, direct execution and simulation of a model is termed
\emph{play-out}. This is achieved by consulting all constraints of the various scenarios before each and every event selection. Thus, the complexity of every runtime decision depends largely only on the number of scenarios, and can ignore the number of states in each scenario, and nondeterministic branching in future system and environment behavior.
By contrast, general program synthesis
approaches for reactive systems (see, e.g., ~\cite{Bloem2012SynthGR1}) apply model-checking, planning, and other techniques toward resolving all specification constraints and environment assumptions a-priori. 
This produces a strategy (e.g., a deterministic finite automaton) for successfully handling all possible environment behaviors at all reachable program states. Synthesis has been applied on SBP specifications with request-and-block idioms in, e.g.,~\cite{Segall2012Synth}. 

One approach for dealing with the large size of state graphs that general synthesis has to analyze, is via run-time planning (also termed on-the-fly/online synthesis) (see, e.g.,~\cite{blum1997graphplan}). In run-time planning or synthesis, the execution mechanism considers a single starting state of the system and the environment, and limits the number of system and/or environment actions in the depth/horizon of the search. This was implemented in SBP in, e.g., \emph{smart play-out} \cite{harel2002smart}. An interesting future research avenue is performing run-time look-ahead or development-time total program synthesis for SBP specifications containing rich constraints specification like the ones shown in this paper. Such research may include identifying categories of constraint specifications that are richer than lists and filters, but are more amenable to synthesis than arbitrary constraints.

Our use of constraint solvers in direct control of executing SBP specifications, is different from other uses of these tools in analysis and verification of systems, including \emph{bounded model-checking}~\cite{BiCiClZh99}, \emph{symbolic execution}~\cite{PaVi09}, \emph{concolic testing}~\cite{Sen07}, and others.
SMT solvers have  been applied in such analysis tasks also in the context of SBP; e.g., by enhancing SMT solvers to deal more efficiently with transition systems ~\cite{KaBaHa15} and by using SMT solvers to efficiently prove compositional properties of a collection of SBP scenarios~\cite{HaKaKaMaMiWe13}. 

\vspace{-0.5cm}

\section{\uppercase{Conclusion}}
\label{sec:conclusion}
\noindent

\vspace{-0.3cm}

Scenario-based programming is a promising approach for the design and
modeling of complex systems, and yet its applicability is somewhat
hindered by the simplistic way in which it interleaves scenario objects. We
proposed here a generalization of the approach that lets objects
interact in much more subtle and intricate ways, and consequently
allows SBP to faithfully model more complex systems. Our
generalization relies heavily on the use of automated constraint
solvers --- tools that are capable of resolving the constraints
imposed by the various scenarios and produce a cohesive
behavior. Apart from setting the theoretical foundations for this
extension, we developed a proof-of-concept implementation and used it
to demonstrate the applicability of our approach.

In the future, we plan to continue this line of work by developing
support for model-checking, statistical analysis and synthesis
algorithms for our extended SBP. These tools exist already for
traditional SBP, and have proven useful --- but extending them
to our formulation will entail accounting for the more flexible 
event selection mechanism. We also intend to apply our extended SBP to
additional, larger case-studies.

\medskip\noindent \textbf{Acknowledgements.}  The work of Assaf Marron
was supported in part by grants to David Harel from the Israel Science
Foundation and the Estate of Emile Mimran.

 \bibliographystyle{apalike}
 {\small

}

\end{document}